\RequirePackage{fix-cm}
\documentclass[twocolumn]{svjour3}          
\smartqed  
\usepackage{graphicx}
\begin{document}

\title{I-band-like non-dispersive inter-shell interaction induced Raman lines in the D band region of double-walled carbon nanotubes
}
\subtitle{}


\author{B\'{a}lint Gyimesi \and
        J\'{a}nos Koltai \and
        Viktor Z\'{o}lyomi \and
        Jen\H{o} K\"{u}rti 
}


\institute{B. Gyimesi \and
                J. Koltai \and
                J. K\"{u}rti \at
                Department of Biological Physics, E\"{o}tv\"{o}s Lor\'{a}nd University, 1117 Budapest, Hungary\\
                \email{gyimesi.balint.88@gmail.com}           
                \and
                V. Z\'{o}lyomi \at
                Department of Physics, Lancaster University, Bailrigg, Lancaster LA1 4YB, United Kingdom\\
                \email{v.zolyomi@lancaster.ac.uk}
}

\date{Received: date / Accepted: date}

\maketitle

\begin{abstract}
Non-dispersive, inter-layer interaction induced Raman peaks (I bands) -- in the region of the D band -- have been observed
recently for bilayer graphene, when the two layers were rotated with respect to each other. Here, similar observations
for double-walled carbon nanotubes (DWCNTs) are theoretically predicted. The prediction is based on double resonance theory,
involving non-zone-centered phonons, and the effect of disorder is replaced by interaction between the two tubes. 

\end{abstract}

\section{Introduction}
\label{intro}

Raman spectroscopy is one of the most powerful methods in investigating carbon nanostructures like graphene or carbon nanotubes. 
The first order Raman spectrum consists of only one band, the G band (in-plane stretching-type motion) for single layer graphene (SLG).
In addition to the G band there is another strong band, the RBM (radial breathing mode) for single walled carbon nanotubes (SWCNTs). However,
due to higher order processes there are further bands allowed in the Raman spectrum of these materials. Some of them -- the D as well
as the 2D (with earlier nomenclature G' or D* used for the latter) bands -- may even be stronger in intensity as compared to the first
order bands. 

The origin of the D and 2D bands is revealed by higher order perturbation theory: they arise due to a fulfilled double resonance
condition and involve non-$\Gamma$-point phonon(s) \cite{Thomsen_PRL2000,Kurti_PRB2002,Venez_PRB2011,Zolyomi_pssb2011}. 
In the case of the D and 2D bands the phonon wave vector ($q$) is not zero but connects two different virtual intermediate $k$-points
which are in the neighborhood of two different K points in the electronic Brillouin zone (BZ): intervalley process with TO phonons
around the K-point. For the 2D band two phonons are involved with opposite wave vector (and same energy). For the D band only one of
the two scattering events between these two $k$-points involves a phonon, the other one is a scattering mediated by a defect.
For each fixed $q$ phonon wave vector the Raman amplitude is obtained by integrating appropriate perturbation formulas like 

\begin{equation}
\sum_{k}\frac{M_{rec}(k,-k)M_{def}(k-q)M_{e-ph}(k,q)M_{e-h}(k,-k)}
{\Delta E_1 \cdot \Delta E_2 \cdot \Delta E_3}
\label{ampl}
\end{equation}
over the virtual intermediate states in the $k$ space \cite{Falicov}. 
Here $M_{e-h}$, $M_{e-ph}$, $M_{def}$ and $M_{rec}$ are the matrix elements corresponding to electron-hole excitation, electron-phonon scattering, electron-defect scattering and electron-hole recombination, respectively.
The energy denominators are:

\begin{equation}
\Delta E_1=E_{L}-(E_{e}(k)+E_{h}(-k)) , \nonumber
\end{equation}

\begin{equation}
\Delta E_2=E_{L}-(E_{e}(k-q)+E_{h}(-k)+\hbar\omega(q)), 
\end{equation}

\begin{equation}
\Delta E_3=E_{L}-(E_{e}(k)+E_{h}(-k)+\hbar\omega(q)) .
\end{equation}

\noindent ($E_L$ is the laser excitation energy, and $\hbar\omega(q)$ is the phonon energy.)

The integral is only large if two of the energy denominators become zero simultaneously. This double resonance condition causes the well known dispersion: the shift of the position of the D (and 2D) bands to higher frequencies with increasing laser excitation energy. 
In the case of graphene-like materials -- due to the Dirac-cones in the electron energy dispersion relation and due to the fact
that the frequency of the TO phonons have a minimum at the K-point -- the absolute value of the D band maximum lies in
the $1300-1400$~cm$^{-1}$ range with a slope of the dispersion of $\approx$ 50~cm$^{-1}$/eV.
Both the position and the slope of the dispersion are twice as large for 2D band as compared to the D band.

\begin{figure}
\includegraphics[scale=0.2]{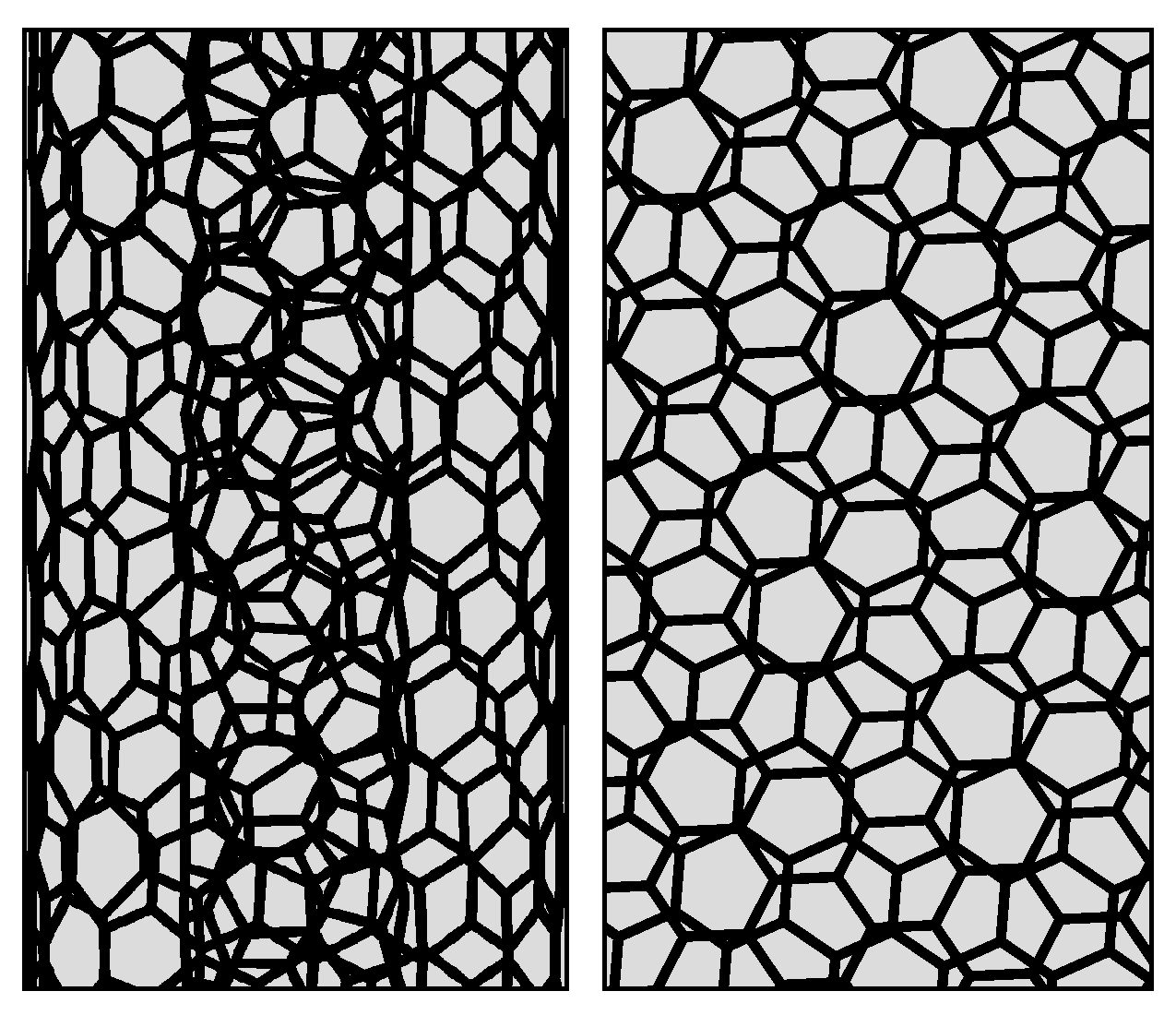}
\caption{Side view of the (4,3)@(14,1) double-walled carbon nanotube, and top view of the same DWCNT, unrolled (BL graphene). Since these structures have a similar pattern, the inter-layer interaction may be comparable.}
\label{fig:0} 
\end{figure}

The mechanism for the D and 2D bands in carbon nanotubes is very similar to that in graphene with one essential difference, namely
the enhancement effect of the Van Hove (VH) singularities has to be taken into account in the case of nanotubes \cite{Kurti_PRB2002,Mault_PRB2013}. 

A different type of peaks in the D band region of the Raman spectrum were observed in bilayer (BL) graphene with rotated layers
recently \cite{Gupta_PRB2010,Righi_PRB2011}. Changing the laser excitation energy changes only the intensity of these bands
but not their position. 
The explanation is based on the interaction between the two layers which can replace the effect of defect scattering.
The role of the interaction can be expressed in two different ways. 
One is to say that one layer imposes a periodic perturbation potential on the other layer with discrete wave vectors which are
the Fourier-components of the Moir\'{e} pattern of the two rotated hexagonal lattices.
The other one is to say that both intermediate scatterings are Umklapp processes, one with the one layer and the other with the
other layer. If both layers have the same orientation, that is, the same reciprocal lattice, nothing happens.
However, if the two reciprocal lattices differ due to the rotation, the difference of two reciprocal vectors does not map onto
the $\Gamma$ point of the first BZ and therefore this nonzero wave vector has to be taken into account in the (quasi)momentum
conservation. In this case there is no continuum for the possible $q$ vectors but they are discrete. Therefore the positions of
so induced lines in the Raman spectrum are fixed, only their intensity, depending on the strenght of the 
double resonance, changes with changing the laser excitation energy. That is, there is no dispersion for
these inter-layer interaction induced Raman lines and of course there is no 2D overtone of them, either.

We show in this work that a similar effect can occur in double-walled carbon nanotubes (DWCNTs) if special conditions
are fulfilled. The interaction between the inner and outer tubes is similar to the interaction between the two layers in BL graphene
and the effect of rotation between the two layers in BL graphene can be replaced by the difference between the chiral angles of the
inner and outer tubes.
Our work was motivated by recent experimental observations where new non-dispersive lines have been observed in the Raman spectrum
of ferrocene-filled carbon nanotubes in the region of the D band, without overtones in the region of the 2D
band \cite{Plank_ACSNano2010,Liu_pssb2011}. One stronger line at 1247~cm$^{-1}$ (C$_1$) and two weaker lines at
1273~cm$^{-1}$ (C$_2$) and 1361~cm$^{-1}$ (C$_3$) were observed \cite{Kuzm_privcom}.

In this paper we show theoretically that the interaction between the inner and outer tubes can indeed give rise to the appearance
of non-dispersive Raman lines in the D band regime.

\section{Theoretical method}
\label{method}

The theoretical procedure can be divided into the following three parts.
\par First, the appropriate wave vector in reciprocal space must be found (connecting the two points determined by each umklapp process).
For simplicity we neglect curvature effects for the electronic states. Instead, we use the two dimensional description of the
states and processes. The one dimensionality is taken into account only by the well known fact that the allowed states form a set
of parallel straight lines, according to the quantization of the component of the wave vectors perpendicular to the tube axis.
The reciprocal lattices, both for the inner and for the outer tubes can be mapped onto a common two-dimensional reciprocal space.
However, they are rotated with respect to each other according to the difference between the chiral angles of the two tubes.
The possible scattering wave vectors can be obtained by the differences between two reciprocal points, one from each of the reciprocal
lattices. However, not all possible differences are allowed, only those for which the
perpendicular component fulfilles the quantization condition. This can only occur if a reciprocal lattice point of one tube fits
exactly onto one of the allowed parallel lines of the other nanotube.
\par Second, the possible wave vectors obtained in the first step are compared to vectors connecting Van Hove singularities in the
two-dimensional $k$ space ($\mathbf{q}$$_{VH}$). The reason for this is the following. The intensity of the I band depends on the
strength of the
double resonance. Here, only the effect of the VH enhancement in the intensity is
taken into account, similarly to the first quantitative description of the usual D band for carbon nanotubes \cite{Kurti_PRB2002}.
The positions of the VH singularities in the two-dimensional $k$ space were obtained using tight-binding (TB) approximation for
the electronic dispersion relation.
\par Third, the phonon frequency for a wave vector satisfying all the necessary conditions described above has to be determined.
We calculate the phonon dispersion relations using density functional theory (DFT) in the helical
BZ \cite{basiuk,DFT_helical,rusznyak_pssb2009}. 
These are performed using the VASP code in a plane-wave basis set, using a plane-wave cutoff energy of 500~eV. The details of the DFT calculations can be found in \cite{basiuk,DFT_helical,rusznyak_pssb2009}.
The determination of the phonon dispersions are carried out using helical coordinates, therefore, the phonon wave vector from the two-dimensional BZ has to be transformed into its equivalent in the one-dimensional helical BZ.

Further details of the theoretical procedure are explained in the next section as the results are presented.

\section{Results and discussions}
\label{results}
\subsection{Possible phonon wave vectors connecting points of different reciprocal lattices}
\label{reciproc}
\par We use nanotube pairs which differ in their radii approximately by the carbon-carbon Van der Waals
equilibrium distance to ensure suitable conditions for interaction. As an example, we choose the (4,3) and the (14,1) chiral
indices for the inner and outer nanotubes, respectively. 
The reason for choosing (4,3) is the motivating experiment mentioned in the introduction. In the experiment the sample had an
RBM mode at 470~cm$^{-1}$ with a maximum intensity at about 676 nm laser excitation which corresponds to a Van Hove transition
energy of $\approx$1.85~eV \cite{Plank_ACSNano2010,Liu_pssb2011}.
The (4,3) tube more less fulfills both conditions: its calculated RBM frequency is 476~cm$^{-1}$ and its calculated E$_{11}$
transition energy is $\approx$1.7~eV.

Fig. 1 shows the reciprocal lattice points of these nanotubes with the allowed momentum-states ($k$ points) of the inner nanotube,
which form -- according to the quantization condition $2\pi R=\lambda_{t}n$ of the tangential component of the electron wave
number $k_{t}=\frac{2\pi}{\lambda_{t}}$, where $n$ is a non-negative integer -- a set of parallel straight lines with a tilt angle 
(with respect to the $x$ axis, which is parallel with a $\mathrm\Gamma$--K direction) of
\begin{equation}
\phi=\arccos\left(\frac{n_{1}+\frac{n_{2}}{2}}{\sqrt{n_{1}^{2}+n_{1}n_{2}+n_{2}^{2}}}\right)+\frac{\pi}{2},
\end{equation}
where the distance between the neighbouring lines is $1 / R$, and $n_{1}$, $n_{2}$ are positive integers, the chiral indices.
$R$ is the radius of the nanotube, which can be expressed with $n_{1}$ and $n_{2}$ as follows:
\begin{equation}
R=\frac{a_{0}\sqrt{n_{1}^{2}+n_{1}n_{2}+n_{2}^{2}}}{2\pi},
\label{radius}
\end{equation}
where $a_{0}$ is the bond length between the carbon atoms (for simplicity, this quantity hereinafter will be considered as
unity) \cite{Reich_book2000}.
\begin{figure}
\includegraphics[scale=0.08]{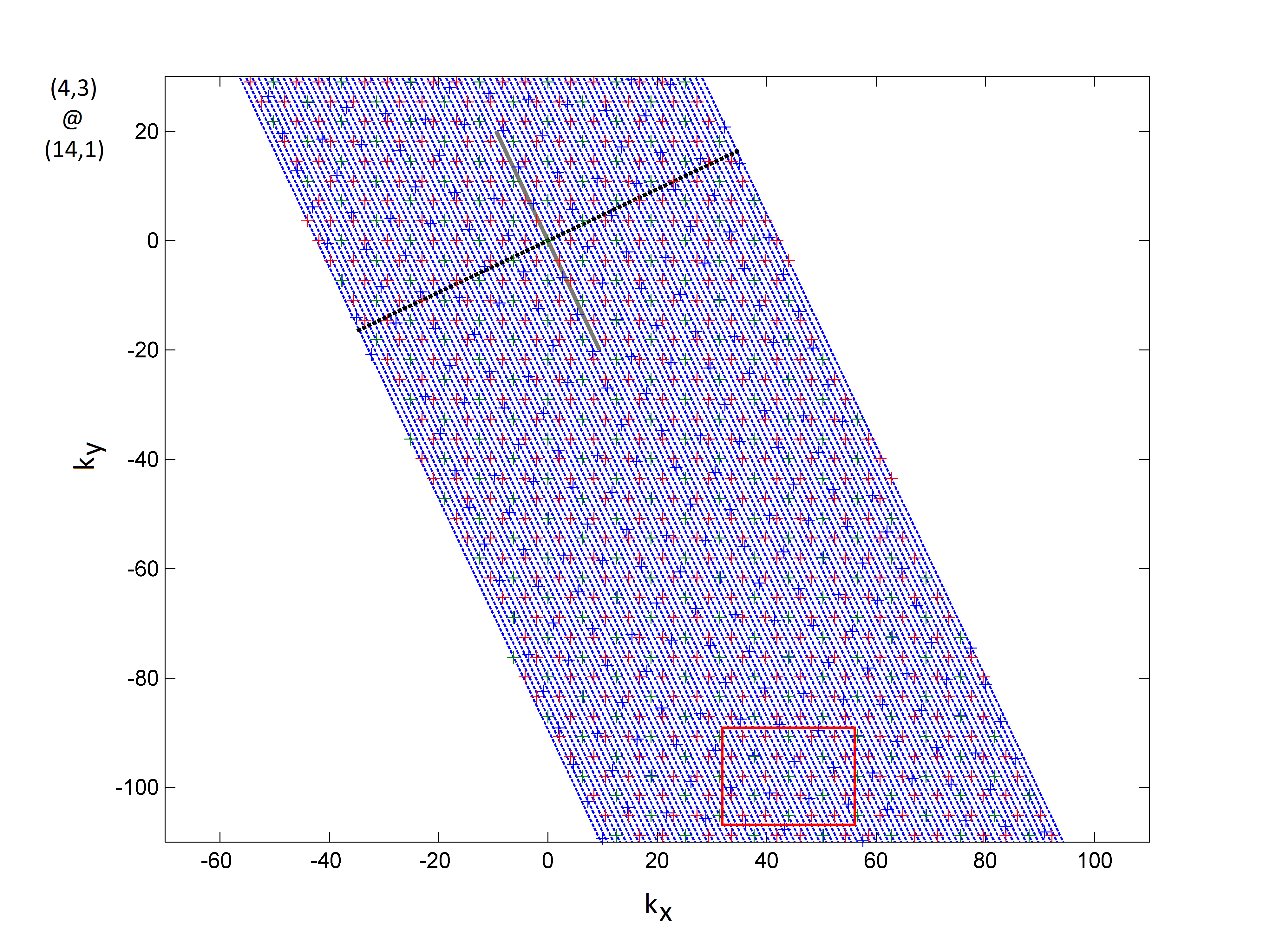}
\caption{The two-dimensional representation of the reciprocal lattices of the (4,3) inner and (14,1) outer nanotubes,
the vertices of the two-dimensional Brillouin zone (BZ) of the inner nanotube, and the allowed momentum states of the (4,3)
nanotube, marked by green, blue and red crosses, and blue dashed lines, respectively. The short black lines and the long grey
line are the one-dimensional first BZs of the inner nanotube in linear and helical representation, respectively. 
}
\label{fig:1}       
\end{figure}
At the bottom of Fig. \ref{fig:1} the red rectangle indicates the surroundings of a reciprocal lattice point of the {\it outer}
nanotube that happens to fit one of the straight lines. Note, that the {\it inner} nanotube's reciprocal lattice points satisfy
this condition automatically. This criterion is the first that must be examined, in order to explain the I band with an interaction
process which is connected to the relative chirality of the two shells in the DWCNT. 
\par
\begin{figure}
\includegraphics[scale=0.08]{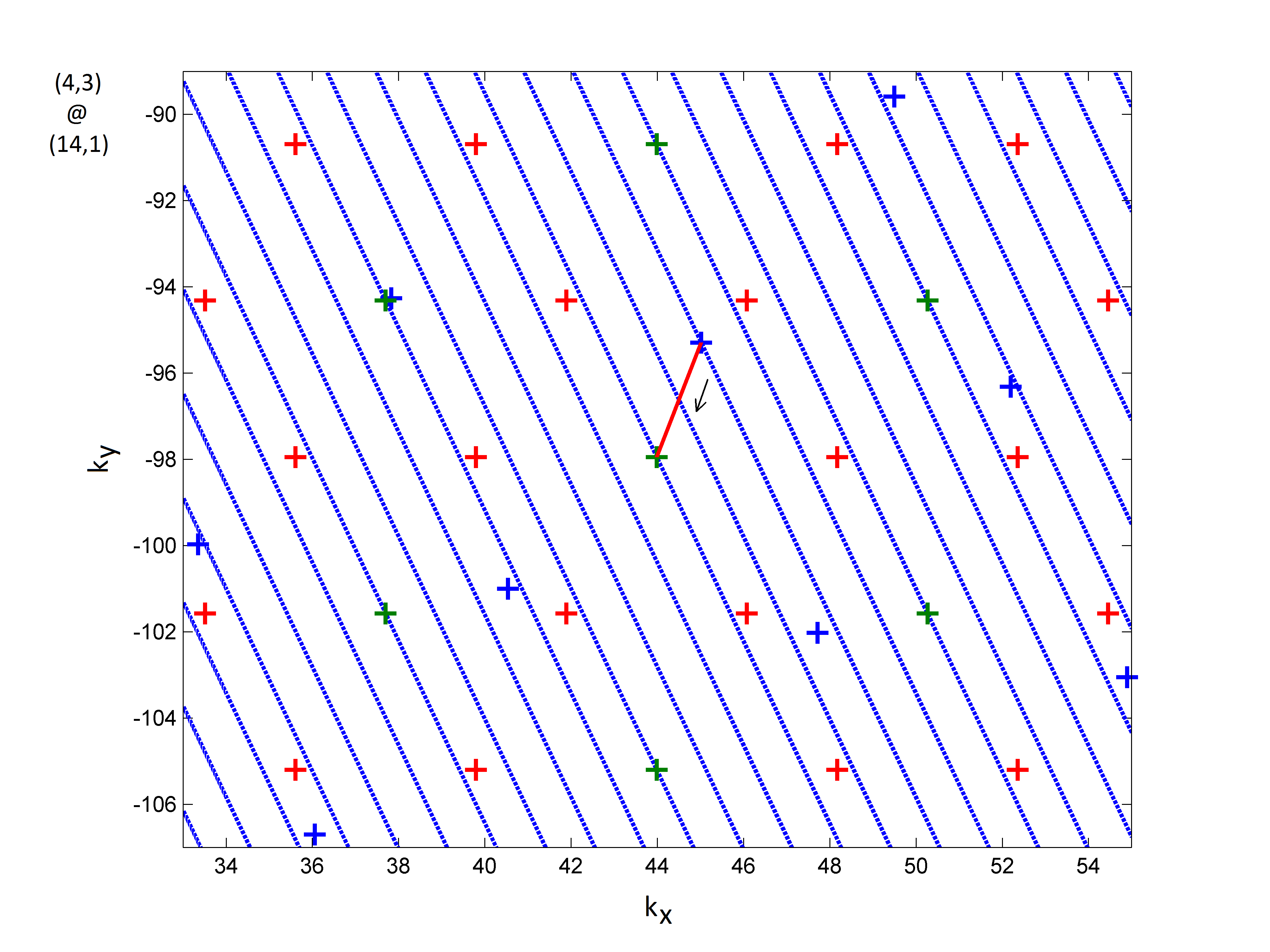}
\caption{A reciprocal lattice point - and its narrow surroundings - of the outer nanotube which is precisely matched to the
inner nanotube's momentum states (the red highlighted section of Fig. \ref{fig:1}). We showed the phonon vector that connects
this point to the nearest reciprocal lattice point of the inner nanotube. The orientation of this vector is also included,
and marked with a small black arrow. Regarding to the notations, the same convention will be used later.}
\label{fig:2} 
\end{figure}

 Fig. \ref{fig:2} is the enlarged version of the previous figure's red framed section, showing a phonon wave vector which can
provide potential contributions to the I band. This vector connects such reciprocal lattice points of the inner and outer nanotubes
that are in accordance with two important requirements. Firstly, the selection conditions of reciprocal lattice points of the outer
nanotube, mentioned above. 
Secondly, due to the quasi-equivalence of the inner nanotube's reciprocal lattice points, that differ only by reciprocal lattice vectors, it is enough to find the reciprocal lattice point of the inner nanotube which is nearest to the given reciprocal lattice point of the outer nanotube. 

We would like to draw attention to another interesting matter. We found that reciprocal lattice points that belong to the outer
nanotube, if they satisfy the criteria mentioned above, are always located on the center line (that crosses the origin). Since
the nanotubes in the DWCNT are concentric, they have coincident axials, and the slope of the parallel lines -- i.e. the allowed
momentum states -- is identical, too, regardless of chirality. Furthermore, due to the different diameter values, these are
following each other with different $1/R$
periodicities that are incommensurate (see Eq. \ref{radius}), except the trivial cases. This has no effect on the centered line. 

It can be shown that the appropriate reciprocal lattice points of the outer nanotube (lying on the center line) can be obtained by a linear combination of the $\vec{k_1}$ and $\vec{k_2}$ reciprocal lattice vectors of the outer tube with coefficients which are integer multiples of the $n_1$ and $n_2$ chiral indices of the outer tube: $\mathbf{k}_{hit}=c(n_1 \vec{k_1} + n_2 \vec{k_2})$.
An interesting fact should be mentioned, namely, for a given inner nanotube, when it is only the parity of the outer tube that deviates (i.e. the case of left-handed and right-handed outer nanotubes), the reciprocal lattices of the outer tubes are mirror images of each other with respect to the center line. Therefore $\mathbf{k}_{hit}$-s remain the same, hence there will be no difference regarding to the phonon wave vectors.

\begin{figure}
\includegraphics[scale=0.08]{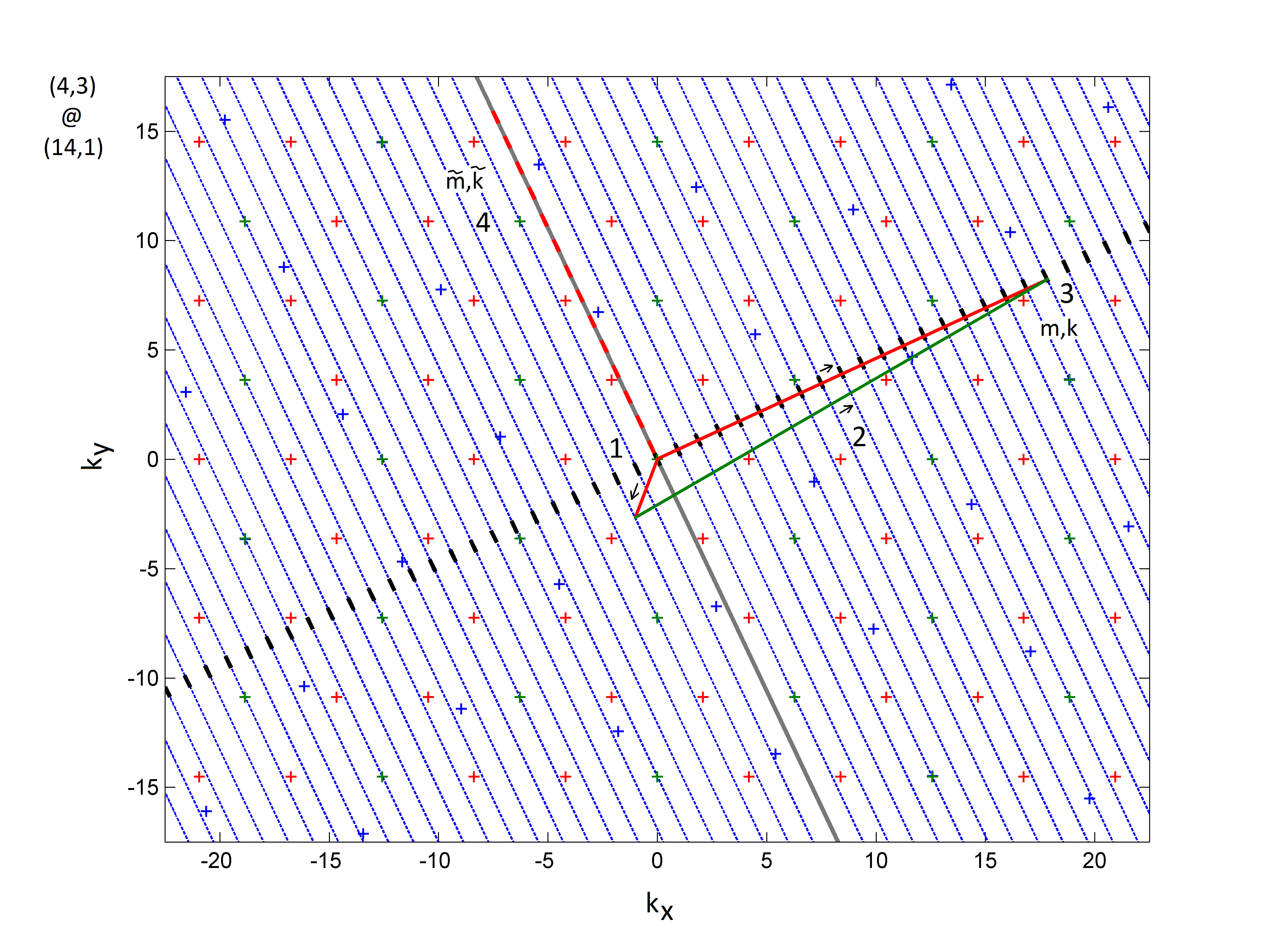}
\caption{Transformations of a given phonon wave vector (marked by red lines) to the origin of the 2D BZ (1), and to the
linear (3) and helical (4) 1D BZs with a simple shift, a reciprocal lattice vector (2, marked by a
green line), and the transformation rules described in the main text, respectively. 
The linear and helical quantum numbers are referred by the signs showed next to the vectors 3 and 4.}
\label{fig:3} 
\end{figure}
\par Beyond the reciprocal lattice points of the outer nanotube that are lying on the center line, searching for other
appropriate reciprocal lattice points is a much more difficult task since we have no analytical formula
that could predict the positions of these reciprocal lattice points as a function of the chiral indices.
However, the number of such points is very low (or even zero), because of the high probability of incommensurability,
according to the lines representing the allowed momentum states. 
\par The next step in searching for the positions of the lines of the I band is to determine the frequencies corresponding
to the phonon wave vectors. Our goal is to obtain the helical quantum numbers of a wave vector so that they can be used for
the one-dimensional helical phonon dispersion calculated using density functional theory in the helical Brillouin
zone \cite{basiuk,DFT_helical,rusznyak_pssb2009}.
The procedure can be viewed on Fig. \ref{fig:3}. We perform transformations that are necessary to determine the linear
and finally the helical quantum numbers which are connected to a given phonon wave vector \cite{Dobar_PRB2003}. First of all, we
move the phonon wave vector shown in Fig. \ref{fig:2} to the origin, into the 1st BZ of the two-dimensional reciprocal space (1). Then - by adding a suitable reciprocal lattice vector (2) - we map the vector into the one-dimensional linear BZ (that is onto one of the
74 short lines of the momentum space, in the case of the (4,3) tube), in order to get the appropriate quantum numbers $m$ and $k$ (3) and from these the helical ones $\tilde{m}$ and $\tilde{k}$ (4).
The linear and helical quantum numbers can be transformed into each other by the following equations:
\begin{equation}
\left|\tilde{k},\tilde{m}\right\rangle=\left|k + \frac{wm}{n}\frac{2\pi}{a}+\tilde{K},\frac{2\tilde{\pi}}{a}m+\tilde{M}n\right\rangle,
\end{equation}
\begin{equation}
\left|k,m\right\rangle=\left|\tilde{k}-\frac{w\tilde{m}}{n}\frac{2\pi}{a}+K\frac{2\pi}{a},\tilde{m}-Kp+Mq\right\rangle
\end{equation}
where $\tilde{K}$, $\tilde{M}$, $K$ and $M$ are integers determined by the condition that the quantum numbers are in the intervals
\begin{equation}
k\in\left(-\frac{\pi}{a},\frac{\pi}{a}\right],m\in\left(-\frac{q}{2},\frac{q}{2}\right],
\end{equation}
\begin{equation}
\tilde{k}\in\left(-\frac{q}{n}\frac{\pi}{a},\frac{q}{n}\frac{\pi}{a}\right],\tilde{m}\in\left(-\frac{n}{2},\frac{n}{2}\right].
\end{equation}
In these equations, $\tilde{\pi}=\frac{\pi q}{n}$, $q$ is the number of graphene lattice points in the nanotube unit cell,
$n$ is the number of lattice points on the chiral vector (the greatest common divisor, g.c.d., of the components of the
chiral vector), $a$ is the translational period of the nanotube, while parameters $w$ and $p$ are non-trivial functions of the chiral
indices \cite{Damnja_PRB1999,Reich_book2000}.

\subsection{The role of Van Hove singularities}
\label{VanHove}
\begin{figure}
\includegraphics[scale=0.08]{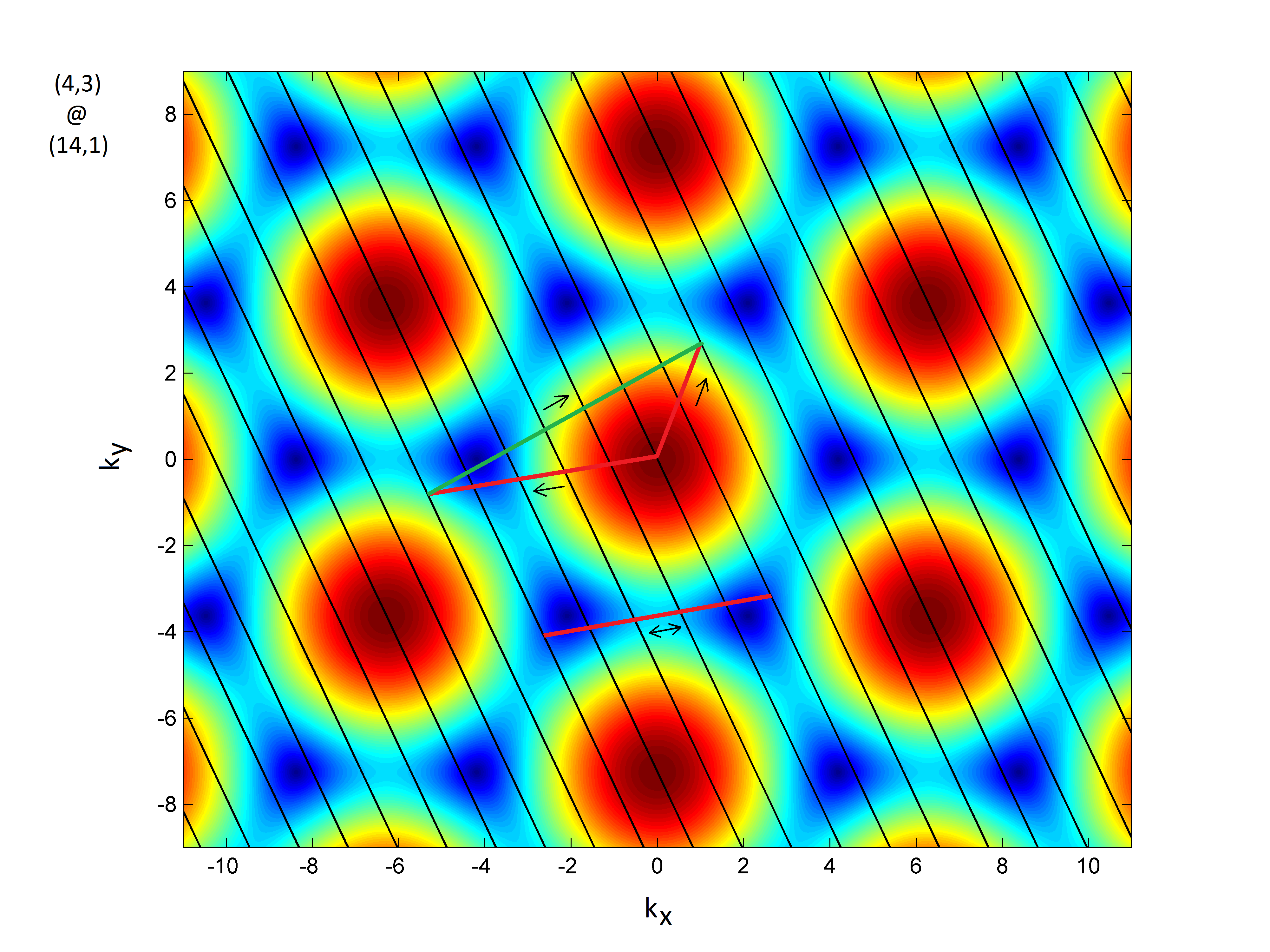}
\caption{Tight-binding electron dispersion relation
$t_{0}\sqrt{1+4\cos{\frac{k_{x}}{2}}\left(\cos{\frac{k_{x}}{2}}\cos{\frac{\sqrt{3}k_{y}}{2}}\right)}$,
where $t_{0} = 2.9$ eV, and $k_{x}$, $k_{y}$ are the wave vector components corresponding to the direction of the axes,
respectively, the allowed states of the momentum space of the (4,3) nanotube (solid black lines), and a transformed
transition vector ($\mathbf{q}$$_{VH}$) which connects two Van Hove singularities. (Note, that there are multiple VH singularities,
but for the sake of simplicity, we focus only on one of them.)}
\label{fig:4} 
\end{figure}
\par There is one important difference between two-dimensional graphene and one-dimensional nanotubes: the existence of
Van Hove singularities in the density of states in the latter case. This has an important consequence: there is an extra
enhancement in intensity if the phonon wave vector connects two $k$ points so that both of them correspond to a VH
singularity \cite{Kurti_PRB2002}. Remaining in the two-dimensional picture: the position of the VH singularities in the
two-dimensional BZ can be found where one of the allowed parallel lines touches an equi-energy curve, that is, perpendicular
to the energy gradient. These particular transitions -- according to the double resonance condition -- must connect points
in the reciprocal space that have the same energy (or at most differ only by the energy of one phonon). Fig. \ref{fig:4} shows
such a transition. The number of transition is greatly reduced by the critera that the allowed momentum states of the nanotubes
are limited to the parallel lines, the
perpendicularity condition of the energy gradient applied to these lines, and the symmetry induced equivalence of the reciprocal
space points. Since the Van Hove singularities
have a comparable amplification factor to the double resonance -- regarding to the intensity -- it is enough to use a similarity
measure for a given phonon vector (that may give a contribution to the I band) that is based on these aspects. In other words,
only the phonon wave vectors connecting a reciprocal lattice point of the inner tube with a reciprocal lattice point of the
outer tube, as discussed in the previous subsection, can give rise to large enough intensity, which matches or almost matches a
$\mathbf{q}$$_{VH}$ wave vector between two VH singularity points.
Accordingly, we define the following phenomenological quantity:
\begin{equation}
C_{I}=e^{-\frac{\left(\mathbf{q}-\mathbf{q}_{VH}\right)^2}{c^2\left|\mathbf{q_{0}}\right|^2}} 
\label{C}
\end{equation}
where $\mathbf{q}$ and $\mathbf{q_{0}}$ belongs to the actual transition and to the vector which connects the inequivalent
K and K' trigonal points, respectively. This formula brings information about the magnitude of the contributions to the I band.
The value of the constant $c$ -- which is a kind of normalization factor -- has been set to $0.1$. In order to perform the
comparison, we made the shift to the origin again, followed by a transformation with a reciprocal lattice vector to move the
Van Hove vector to the first Brillouin zone (see Fig. \ref{fig:4}).
Note, that the determination of the positions of the VH singularities was carried out without taking
into account curvature effects (relying on the two-dimensional tight-binding approximation for electron energies combined
with the discrete parallel lines for the allowed wave vectors.). Of course, considering $\sigma$-$\pi$ rehybridization due
to the curvature would influence strongly the electron energies \cite{Zolyomi_PhD} and therefore also $\mathbf{q}$$_{VH}$ if
the nanotubes have small diameters.

\subsection{I band frequencies}
\label{freq}
\begin{figure}
\includegraphics[scale=0.073]{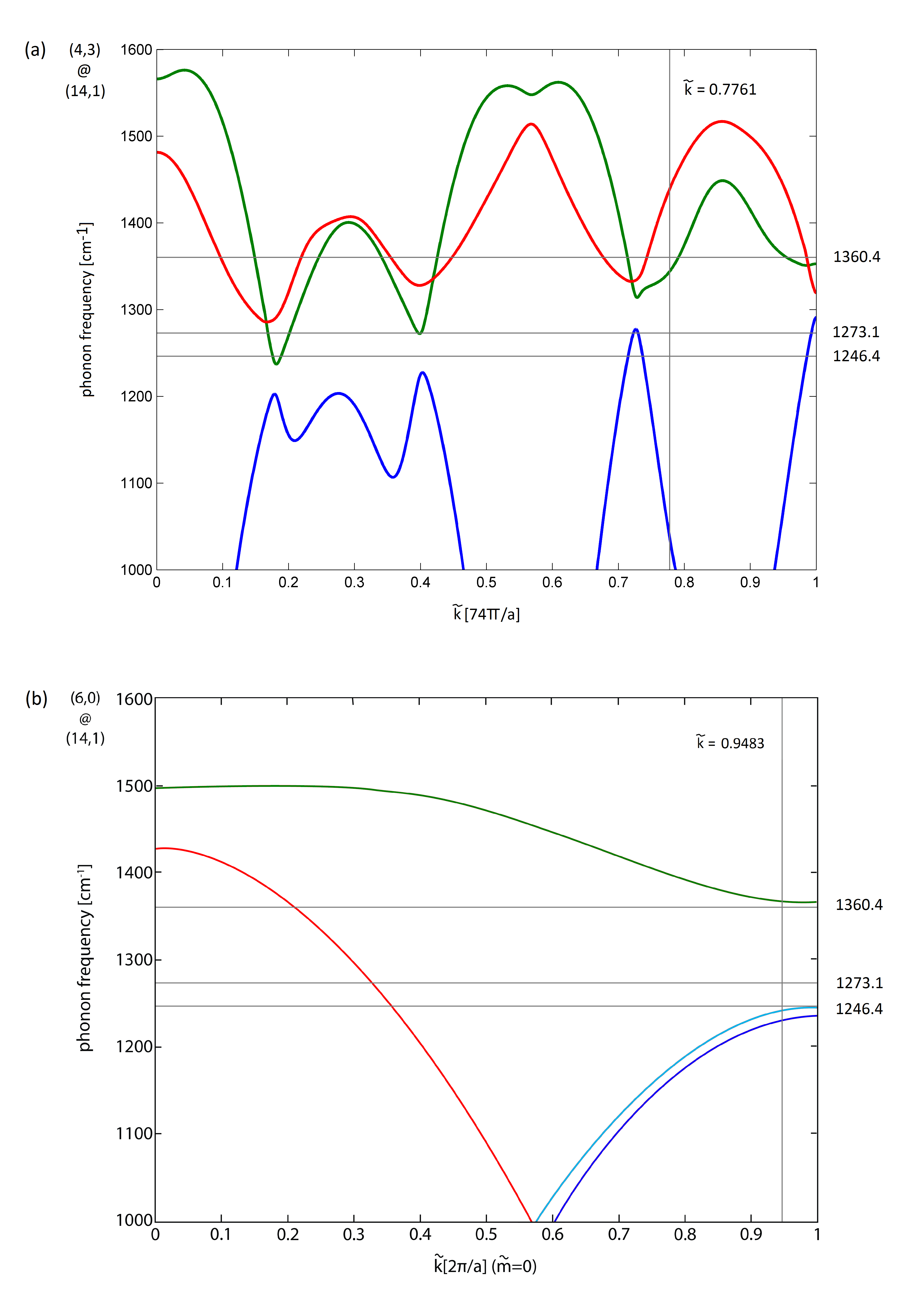}
\caption{Phonon dispersions of the (4,3) (a) and (6,0) (b) nanotube (calculated via DFT), which are characterized by the
$\tilde{m}=0$ helical quantum number. The vertical grey lines represent the helical wave numbers that have been
specified in the reciprocal space of the (4,3) and (6,0) inner and (14,1) outer nanotubes.}
\label{fig:5} 
\end{figure}
\begin{figure}
\includegraphics[scale=0.4]{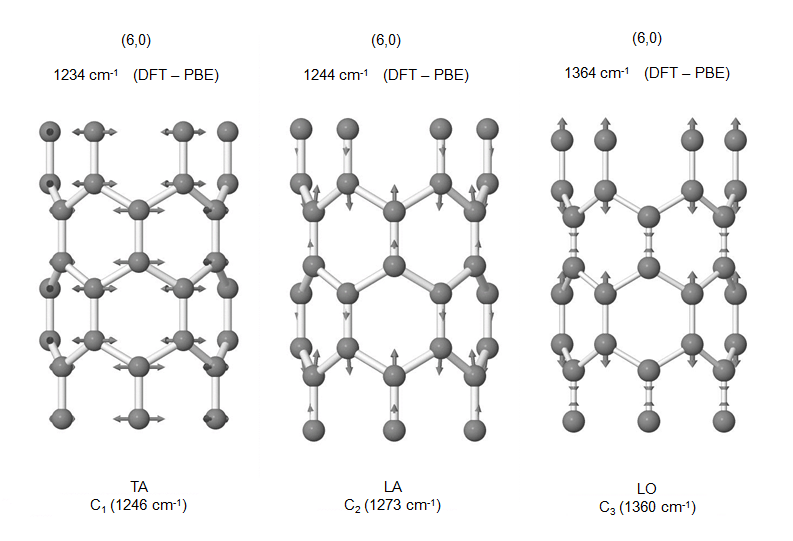}
\caption{Certain vibrations of the (6,0) nanotube,
calculated via DFT Perdew-Burke-Ernzerhof model. 
The TA, LA and LO abbreviations correspond to transverse
and longitudinal acoustic and optical modes, respectively. 
}
\label{fig:6} 
\end{figure}
After finding the possible combinations of reciprocal vectors of the inner and outer tubes one has to compare them with
the $\mathbf{q}$$_{VH}$ Van Hove vector and consider only those $q$ values for which the $C_I$ quantity as defined in Eq. \ref{C} is close
to unity. All these conditions together are rather strict so the number of the wave vector candidates for the I band is small.
In fact, for the case of (4,3)@(14,1) we found only one such wave vector. After the appropriate transformations (as shown in Fig. 3)
we obtained $\tilde{k}=0.7761$ for the helical wave number of this vector. 
(The $\tilde{m}$ quantum number is zero, because the g.c.d. is one for (4,3).) For comparison, the quantum number for the Van Hove
vector is $\tilde{k}_{VH}=0.7762$ for (4,3).

Fig. \ref{fig:5}a shows the DFT calculated phonon dispersion of (4,3) nanotube in the helical BZ. The vertical grey line shows the
position of the $\tilde{k}=0.7761$ helical wave number. This line intersects three phonon dispersion curves. 
The three horizontal grey lines represent the measured frequencies mentioned in the Introduction, for comparison.
It can be seen that although one of the obtained frequencies matches quite well the measured 1360~cm$^{-1}$, the other
two calculated values are far away from the measured ones.

We repeated the whole procedure for another combination: (6,0)@(14,1). The diameter as well as one of the Van Hove excitation
energies of the (6,0) nanotube is very similar to those of the (4,3) tube (see Fig. 6 in Ref. \cite{Plank_ACSNano2010}). 
For the zig-zag (6,0) nanotube it is easy to show that the $\tilde{k}_{VH}$ vector connecting two VH singularities
has $m=6$ and $k=0$ linear quantum numbers which transforms into $\tilde{m}=0$ and $\tilde{k}=1$ helical quantum numbers. 
We found a $q$ vector between the reciprocal points of the inner and outer tubes which is very close to the aforementioned
$\tilde{k}_{VH}$ vector. Its helical wave number is $\tilde{k}=0.9483$.

Fig. \ref{fig:5}b shows the DFT calculated phonon dispersion of (6,0) nanotube in the helical BZ (for $\tilde{m}=0$). The
vertical grey line shows the position of the $\tilde{k}=0.9483$ helical wave number. The three horizontal grey lines represent
again the measured frequencies mentioned in the Introduction. It can be seen that the so obtained frequencies are quite close
to the measured ones.

Fig. \ref{fig:6}a shows the vibrational pattern for these three normal modes. The two with lower frequencies would be degenerate
in the planar case. The curvature results in a small splitting of their frequencies.

\section{Conclusions}
\label{sum}

In conclusion, we have shown that similar to the appearance of the non-dispersive Raman D band activated by well-ordered
inter-layer interactions in rotationally stacked bilayer graphene (I band) the interaction between the inner and outer
tubes with different chiralities can result in the appearance of non-dispersive I band in double-walled carbon nanotubes as well. 

We have shown on the examples of (4,3)@(14,1) and (6,0)@(14,1) DWCNTs that the strict conditions for appropriate wave vectors
for inducing the non-dispersive I band in the D band region can be fulfilled. We focused on the most important factor, namely that
the difference between two reciprocal lattice points -- one for the inner and one for the outer tube -- matches or nearly matches a
vector connecting two VH singularities in $k$ space. 
Of course the double resonance condition is only fulfilled for the laser excitation energy which matches the VH transition energy.
Interestingly, the enantiomers can not be distinguished, because the interaction which results in the I band of a given inner tube with the left handed and with the right handed form of the outer tube of the same chirality turned out to be exactly the same.

There are further factors which influence the intensity of the I band, namely the matrix elements, of which the most important
is the electron-phonon coupling constant. This will however only be important for the exact intensity of the non-dispersive line;
neglecting the matrix elements therefore does not impact the message of this work, namely that the appearance of I-band-like lines is
possible in DWCNTs.

\begin{acknowledgements}
Support from the Hungarian National Research Fund (OTKA), grant numbers K81492 and K108676 are gratefully acknowledged.

\end{acknowledgements}



\end{document}